\documentclass[aps,prl,twocolumn]{revtex4-1}
\usepackage[latin1]{inputenc}
\usepackage{amsmath}
\usepackage{amsfonts}
\usepackage{amssymb}
\usepackage{bm}
\usepackage{array}
\usepackage{color}
\usepackage{graphicx}
\usepackage{slashbox}
\usepackage{multirow}

\begin{document}

\title{Augmented collisional ionization via excited states in XUV cluster interactions}


\author{Edward Ackad, Nicolas Bigaouette and Lora Ramunno}
% \email[]{lramunno@uottawa.ca}
% \homepage[]{www.science.uottawa.ca/lramunno}
%\thanks{}
\affiliation{Department of Physics, University of Ottawa, Ottawa, Ontario K1N 6N5, Canada}


\date{\today}

\begin{abstract}

The impact of atomic excited states is investigated via a detailed model of laser-cluster interactions, which is applied to rare gas clusters in intense femtosecond pulses in the extreme ultraviolet (XUV).  This demonstrates the potential for a two-step ionization process in laser-cluster interactions, with the resulting intermediate excited states allowing for the creation of high charge states and the rapid dissemination of laser pulse energy.  The consequences of this excitation mechanism are demonstrated through simulations of recent experiments in argon clusters interacting with XUV radiation, in which this two-step process is shown to play a primary role; this is consistent with our hypothesis that XUV-cluster interactions provide a unique window into the role of excited atomic states due to the relative lack of photoionization and laser field-driven phenomena.  Our analysis suggests that atomic excited states may play an important role in interactions of intense radiation with materials in a variety of wavelength regimes, including potential implications for proposed studies of single molecule imaging  with intense X-rays.
\end{abstract}

\pacs{36.40.Wa,34.80.Dp,52.65.-y}

\maketitle

% INTRODUCTION

Over the last decade, coherent radiation sources have been developed that can probe light-matter interactions at ever smaller wavelengths and unprecedented intensities \cite{wabnitz_nature,lcls_nature}. 
Recent experiments have moved into the extreme ultraviolet (XUV), including several that have explored intense XUV interactions with rare gas clusters \cite{PhysRevLett.100.133401,murphy:203401}. 
The XUV regime near 30 nm is unique because the photon energy is too small for inner shell ionization of rare gas atoms, yet too large (at reported intensities) for any appreciable laser field-driven processes that dominate intense laser-cluster interactions at longer wavelengths, such as collisional heating. We therefore propose that
this regime presents a unique opportunity to isolate the influence of the internal electronic structure of atoms within clusters on intense radiation-cluster interactions.
 
To date, models of cluster interactions with intense radiation in the near IR \cite{RevModPhys.82.1793} or in the vacuum ultraviolet (VUV) \cite{PhysRevLett.93.043402,PhysRevLett.91.233401,PhysRevLett.99.233401,mbr} have not incorporated the effects of the atomic structure of the constituent atoms.
 As a result, the influence of the atomic structure in such interactions is unknown over a broad range of wavelengths, including in the XUV. 

In this Letter, we introduce a general model that explicitly incorporates the effect of atomic and ionic excited states on collisional ionization, and apply it to a molecular dynamics code for rare gas cluster interactions with intense radiation.
By including the process whereby a ground-state electron of any charge species can be internally promoted to an excited state by a colliding electron, ionization is allowed to occur through a two-step process: excitation followed by ionization from the excited state.  
This requires two sequential collisions, each of less energy than is required by single step collisional ionization from the ground state (the conventional model used widely across wavelength regimes). This generally applicable two-step process, which we call ``augmented collisional ionization" (ACI), is applied to the 32 nm experiment with Argon clusters \cite{PhysRevLett.100.133401}. ACI is found to dominate this interaction, and allows the system to access higher ionic charge states than are obtained via the single step collisional ionization process alone. 

The incorporation of ACI into our laser-cluster interaction model provides the first explanation of the observation of high charge states (Ar 4+) in the 32nm system. Further, we find that cluster atoms quickly become excited and then ionized through ACI, demonstrating that the pulse energy is spread rapidly through the cluster.  This work therefore provides a fundamental insight into the dynamics of XUV-cluster interactions, an area that has, to date, been the subject of only a small number of theoretical investigations \cite{PhysRevLett.100.133401,murphy:203401,ZiajaVUV_XUV}. 
Several of these focused on the electron emission spectra of the 32 nm - Argon interaction, though did not account for the experimentally observed high charge state.
In particular, a multistep ionization model was proposed that explained the electron spectra for all but the highest pulse intensities \cite{PhysRevLett.100.133401}, whereas 
Ziaja \textit{et al.} obtained good agreement with the emission spectra \cite{ZiajaVUV_XUV} using a Kinetic Boltzmann model including non-equilibrium and equilibrium dynamics.

Through our model, we find that atomic excited states in collisional systems play a crucial role in understanding both the charge state spectrum and how quickly charge states are produced due to increased energy transfer between electrons and ions. 
Though the XUV provides a unique regime in which to study these effects, our results are of more general application. As a result, this work opens the door to the study of the importance of excited states in other material systems and in wavelength regimes ranging from the infrared to X-rays. 
Further, the more rapid energy transfer to ions through the ACI process could also play an important role in explosion dynamics, which may be an important consideration in proposed single molecule imaging studies with intense X-rays \cite{lcls_imaging_nature}.

% METHOD

In this work, the motion of the particles is calculated via classical molecular dynamics. Quantum effects are included by determining transition cross-sections and using a Monte-Carlo scheme to determine when excitation or ionization occurs. In the case of ionization, a 
free electron is added to the dynamics simulation. In the case of excitation, no new electron is added, but the parent atom/ion is set to be in an excited state, which determines its future ionization potential and cross-section.

%photoionization

The photoionization probability is calculated at each time-iteration for every atom/ion, and this is used to test for photoionization events. The photon flux is determined via the intensity, which is modulated both by the time profile of the pulse and by photon absorption. The cross-section for the photoionization of neutral Argon was obtained from Ref.~\cite{Argon_photo_neutral}, while those of ionic Argon were obtained using Los Alamos Atomic Physics Codes \cite{gipper}.

% collisional ionization

The atom-electron collision in the cluster environment is modeled as an isolated atom-electron collision in a background cluster potential. This allows the collisional cross-sections for both the ionization and excitation to be calculated using an isolated atom-electron model. Single-step collisional ionization cross-sections were calculated using the semi-empirical Lotz formula \cite{Lotz}. The Born plane-wave approximation \cite{cowan} was used for both excitation and ionization from an excited state. Computed collisional excitation cross-sections were compared to experimentally measurable excitations to metastable states; our results for neutral Argon closely match experimental results for the metastable part of the $3p^6 \rightarrow 3p^54s$ excitation ($^3P_0$ and $^3P_2$) \cite{PhysRevA.9.1195}.

% excited states (wrt collisional)

The excited states used in this work consist of a subset of all possible excited states; only single electron excited states were used. Of those, only the lowest eight states with $l<4$ were implemented. This subset is the most important, as it contains the lowest energy states. Including more single electron excited states adds to the total collisional cross-section, although states near the threshold require almost as much energy as ionization and are thus almost as infrequent. The energy of the excited states and the cross-sections were obtained using the Hartree-Fock implementation of the Cowan code \cite{cowan}.

We find that the inclusion of excited states in our model does not alter photonionization. This is because photoexcitation away from a resonance is rare as only states of $|\Delta l|=1$ are accessible. At the same time, the photon wavelength must also be approximately equal to the transition energy. In electron-atom collisions, in contrast, the electron can transfer any amount of its orbital angular momentum and its available kinetic energy.

% justification of threshold.

The influence of the cluster is accounted for by determining the potential $V_p$ at the atom due to all particles farther than the nearest neighbor distance, $R$ ($>$ 6.4 Bohr for Argon), and approximating it as a constant external potential in the vicinity of the atom. The potential at the atom is reinterpreted as the cluster-free threshold of the atom, similar to Ref.~\cite{PhysRevLett.99.233401}.

The constant potential approximation is justified by considering an atom at $\vec{r}=0$ that is surrounded by two neighboring ions on the z-axis at $\pm \vec{R}$. The true background potential at the atom is given by
\begin{eqnarray}
\label{pots}
V_p(\vec{r}) &=&
\frac{-Z_1}{|\vec{r}-\vec{R}|}+ \frac{-Z_2}{|\vec{r}+\vec{R}|} \\ &=& -Z_1\sum_{l=0}{\frac{ r_<^l}{r_>^{l+1}} }P_l(\xi)
 -Z_2\sum_{l=0}{\frac{ (-1)^l r_<^l}{r_>^{l+1}} }P_l(\xi),
\end{eqnarray}
where $\xi=\cos(\theta)$, $P_l$ is the Legendre polynomial of order $l$, $Z_1$ and $Z_2$ are the charges of the neighboring ions.

Both neighbors contribute a constant term, $Z/R$, to the Hamiltonian of the atom which merely shifts the threshold.
Using first-order perturbation theory, the lowest order correction to the eigenvalue is,
\begin{equation}
\Delta E =  -\frac{Z_1+Z_2}{R^3}\left\langle \phi_0 \left| \hat{q} \right| \phi_0\right\rangle,
\end{equation}
where $\hat{q}$ is the quadrupole operator. The state $\phi_0$ is an eigenstate of the atomic Hamiltonian, and therefore the angular momentum operator, making the dipole term zero. The potential of the electrons within $R$ of the atom is also removed from $V_p$. The collisional system therefore closely approximates an isolated system.
% Screening
The screening effects of the electrons within $R$ on the ionization potential are not easily modeled microscopically due to the highly collisional nature of the system. Since ACI would not be fundamentally changed by screening effects, they are not included here.

During a collision, the energetically accessible states are determined by the kinetic energy of the impacting electron relative to the target atom's cluster-free threshold. This includes ionization if energetically permissible. The total cross-section for all of the possible final states is used to determine if any excitation or ionization event will occur. If the electron's impact parameter, $b(=|\vec{v} \times \vec{r}|/|\vec{v}|)$, is within this total cross-section, then a state is chosen at random, weighted by its cross-section relative to the total cross-section.

% decay mechanisms for excited states

In our model, we have not included any mechanisms for the decay of excited states. While it is possible for the excited states to decay radiatively, the shortest lived states will be states of $|\Delta l| =1$. These decays are on the order of nanoseconds \cite{bruce:2917,eichhorn:026401} making them unlikely to have a large effect on our results.

Collisional de-excitation, which entails an impact electron gaining energy by de-exciting the atom, is more probable though still unlikely to play a role. This is because the cross-section for ionization from an excited state is much larger than the de-excitation cross-section, favoring ionization provided the impact electron has the small amount of energy needed to ionize. Further, even if we were to allow collisional de-excitation to occur, it would be unlikely to significantly alter our results, since any impact electron that would gain energy would more than likely use that extra energy to excite or ionize another atom, leading to the same results, on average.

% RESULTS

% description of the simulations

To determine the importance of ACI, we simulated the interaction of Argon clusters exposed to 32.8 nm radiation both with and without the ACI mechanism included. Conventional collisional ionization was included in all simulations. 
We used a pulse width of 25 fs, and varied the intensity from  $5\times10^{13}$ to $1\times10^{14}$W/cm$^2$. Though the reported average cluster size in Ref.~\cite{PhysRevLett.100.133401} was 80 atoms, a population of larger clusters is possible in the cluster jet. Thus we considered two sizes: Ar$_{80}$, and the closed-shell icosahedral structure Ar$_{147}$. The smaller closed-shell icosahedral, Ar$_{55}$, may also be present but would likely have a much smaller contribution to the Ar$^{4+}$ signal. For each set of parameters, we performed ~10$^4$  separate simulations, and we ran each Ar$_{80}$ simulation to 800 fs and each Ar$_{147}$ simulation to 1 ps.

% description of what data is listed in table 1

In Table \ref{ept}, we list the average yield of Ar 4+ per cluster for each set of simulations. The number on the left was obtained from the inner-ionized Ar 4+ observed at the end of the simulations ({\it i.e.}, not taking into account the electrons that are loosely bound to individual atoms). The number on the right was obtained by excluding any 4+ ions that have at least one bound electron at the end of the simulation; this then gives a more accurate picture of what would be observed at the detector.

 \begin{table}[b]
 {\small
 \begin{tabular}[b]{|l|l|l|} \hline
    \multicolumn{3}{|c|}{Ar$_{80}$}  \\ \hline
 \backslashbox{Intensity}{ACI}  & No & Yes  \\ \hline
 $5\times10^{13}$ W/cm$^2$ & 0/0 & 0.0114/.0020  \\ \hline
 $7.5\times10^{13}$ W/cm$^2$& 0.0002/0  & 0.060/0.0045   \\ \hline
 $10\times10^{13}$ W/cm$^2$ & 0.0008/0 & 0.149/.0099  \\ \hline
  \multicolumn{3}{|c|}{Ar$_{147}$}  \\ \hline
 & No & Yes \\ \hline
 $5\times10^{13}$ W/cm$^2$ &  0/0 & 0.0338/0.0076   \\ \hline
 $7.5\times10^{13}$ W/cm$^2$ &  0.0008/0  & 0.151/0.0258   \\ \hline
 $10\times10^{13}$ W/cm$^2$ & 0.0037/0  & 0.4183/0.0509 \\ \hline   \end{tabular} }
 \caption{\label{ept} Average yield of Ar$^{4+}$ at different intensities for Argon cluster sizes of 80 and (closed-shell icosahedral) 147 with (left) and without (right) two-step augmented collisional ionization (ACI) mechanism. 
The numbers on the right side were obtained by excluding any 4+ ions that have at least one bound electron at the end of the simulation.
}
 \end{table}

Table~\ref{ept} shows that Ar$^{4+}$ only emerges at the intensity quoted in Ref.~\cite{PhysRevLett.100.133401} when the ACI mechanism is included. Though we see some inner ionized 4+ states without ACI at higher intensities, the yield is orders of magnitude less than when ACI is included, and there were no 4+ that were bare ions, meaning they would be detected as lower charge states. In contrast, when ACI was included, some 4+ always remained without bound electrons and would thus be detected as 4+. 
While we see an increase in the number of 4+ in the larger versus smaller clusters, the influence of cluster size is still smaller than the increase from the ACI ionization channel. Thus we find that, at all intensities, ACI is the single largest effect.

% end of table 1

% figure 1

\begin{figure}[ht]
\includegraphics[angle=0,scale=1]{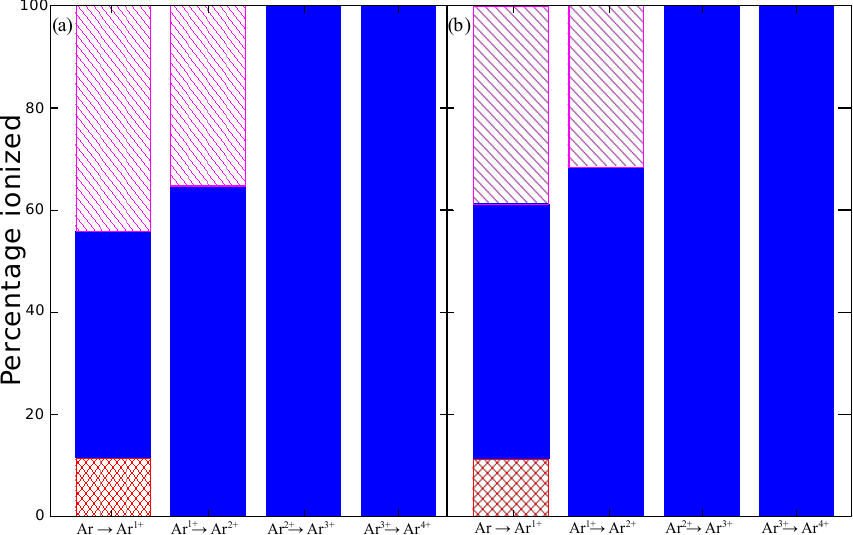}
\caption{(color online) The relative contributions of photoionization (diagonal lines), single step ionization (boxes) and ACI (solid) in the creation of each ion species of Argon by the interaction of (a) Ar$_{80}$ clusters and (b) Ar$_{147}$ clusters with 32.8 nm radiation at an intensity of $5\times10^{13}$ W/cm$^2$.}\label{ionfigs}
\end{figure}

To further illustrate the dominance of ACI, Figure~\ref{ionfigs} plots that fraction of each ionization channel that led to the creation of the various charge states. The left histogram is for Ar$_{80}$ while the right is for Ar$_{147}$.
Photoionization contributes only to the creation of Ar$^{+1}$  and Ar$^{+2}$, as expected.
Since photoionization plays a lesser relative role in the larger clusters, we see that clusters become more collisional as their size increases.
Single step collisional ionization without excited states contributes significantly only to the creation of Ar$^{+1}$. ACI, however, dominates the creation of Ar$^{+2}$ to Ar$^{+4}$, and is the exclusive mechanism for the creation of Ar$^{4+}$.

%%%% Eddie: the shading in black and white is unclear...

% figure 2 main

\begin{figure}[ht]
\includegraphics[angle=0,scale=0.425]{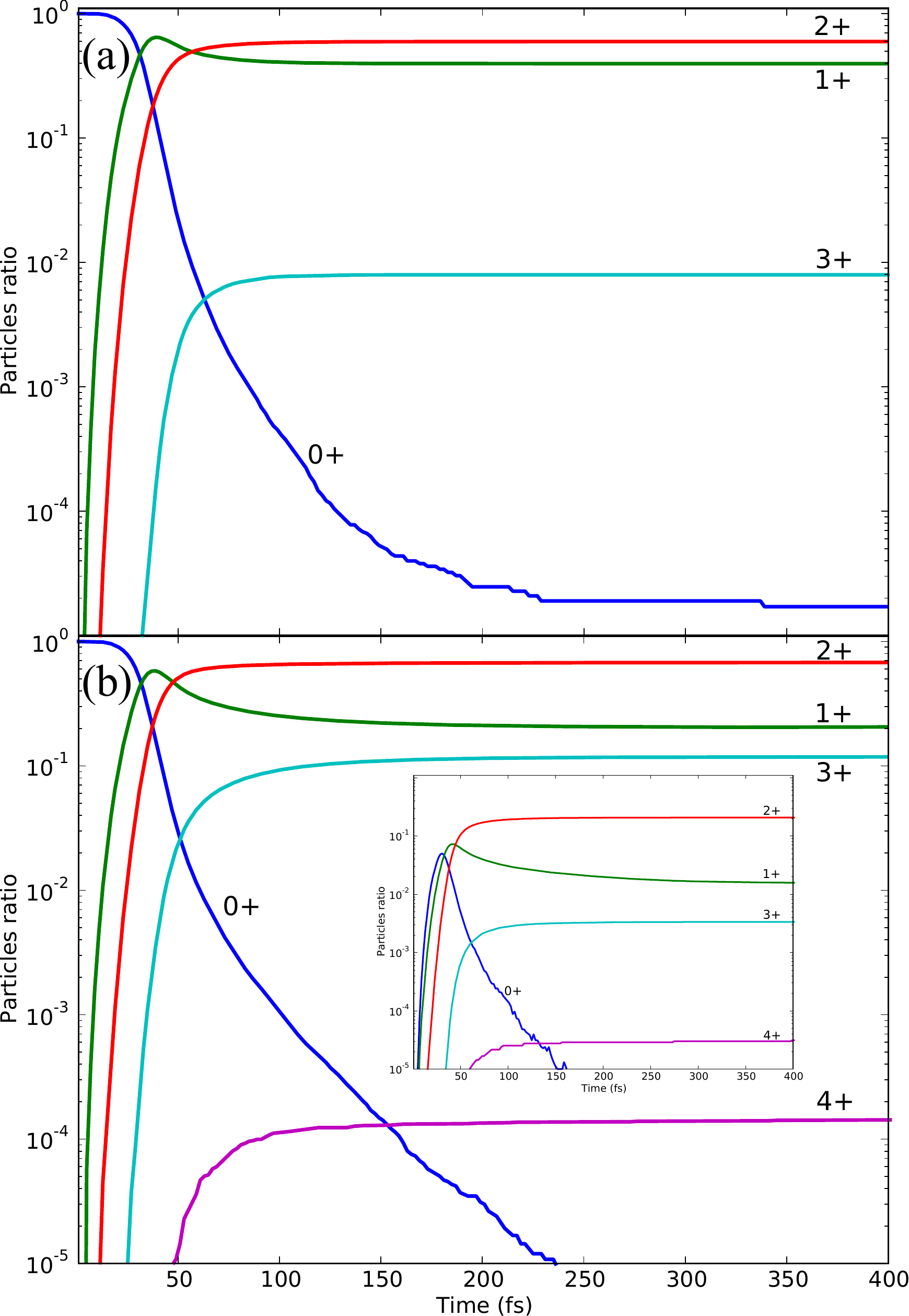}
\caption{(color online) The charge species population relative to the total number of atoms and ions as a function of time for Ar$_{80}$ clusters interacting with radiation of intensity 5$\times10^{13}$ W/cm$^2$, both without (a) and with (b) ACI. The insert in (b) gives the relative excited charge species population as a function of time.}\label{chgev}
\end{figure}

Next, we examine the dynamics of the interaction to further explore the role of ACI and the prevalence of the excited states. Fig.~\ref{chgev} plots the relative (inner ionized) charge state population as a function of time during the interaction for Ar$_{80}$ clusters both without (top) and with (bottom) the ACI mechanism included. Prior to the peak of the pulse (at 34.3 fs), atoms are the predominant species regardless of whether or not ACI is present. Soon after the pulse peak the Ar$^{1+}$ species is the most prevalent. This lasts until about the end of the pulse at 68.6 fs when the Ar$^{2+}$ becomes the most prevalent.

With ACI, the higher charge states appear earlier and in greater abundance. The peak in the Ar$^{1+}$ population happens roughly 2 fs earlier with ACI, the Ar$^{2+}$ surpasses Ar$^{1+}$ around 10 fs earlier, and a sizable Ar$^{3+}$ population emerges 7 fs earlier and surpasses the atomic population around 12 fs earlier than without ACI. The final abundance of Ar$^{3+}$ is more than an order of magnitude larger when ACI is present. In both cases the Ar$^{2+}$ is the most abundant species at the end, but the ratios between species populations are very different. With ACI we also see the emergence of Ar$^{4+}$, which appears at the tail of the laser pulse but reaches a plateau around 100 fs later, at which time it surpasses the abundance of atoms.

% figure 2 inset

The insert in Fig.~\ref{chgev} displays the fraction of each charge species that is in excited states as a function of time. The abundance of the excited states (after the pulse peak) closely follows the abundance of the internal charge states and a sizable fraction of each charge species is excited.

%The abundance of excited atoms/ions peaks around 30 fs while the peak of excited Ar$^{1+}$ is around 41 fs. The excited Ar$^{1+}$ peak coincides with the peak in the abundance of the Ar$^{1+}$. The abundance of the excited states after the peak of the laser pulse follow the closely the abundance of the internal charge states. Thus at any given time after the laser pulse a sizable fraction of each charge species is excited.

ACI thus allows the cluster to reach high charges and to do so earlier than would otherwise be possible. This is a result of having a significant proportion of each charge species excited. Consequently, impact electrons do not require much kinetic energy to ionize the excited electrons as compared to single step ionization or the excitation step itself. For larger clusters, in which more electrons are trapped by the cluster potential, the peak abundance of excited states occurs even earlier than in small clusters at the same intensity (not shown).

% CONCLUSION

In summary, we have shown that the detection of the Ar$^{4+}$ in Ref.~\cite{PhysRevLett.100.133401} can be explained by augmented collisional ionization (ACI), a two-step process wherein collisional ionization occurs via intermediate atomic and ionic excited states.  ACI was found to dominate the cluster-XUV interaction dynamics. In particular, it occurs much more frequently than direct collisional ionization from the ground state, which alone cannot reproduce the high charge state observed in experiment. While we have applied our model to cluster-XUV interactions, ACI will likely play a role in any system that is highly collisional, including in a variety of finite condensed systems interacting with intense radiation in a wide range of wavelength regimes.

We have also shown that ACI provides a route to high charge states in a shorter time, due to the efficacy of the energy transfer by ACI afforded by the lower energy gaps of excitation compared to direct ionization. The energy imparted to a small finite system will thus be highly sensitive to the ramp up of the pulse, which may have consequences for the direct imaging of single biomolecules with short, intense Xray pulses. Thus a good understanding of excited states of the photo-sensitive parts of the biomolecule ({\it i.e.}, atoms and molecules that have largest photoionization cross-sections), and a sharp ramp up of the imaging pulse will be crucial.

\begin{acknowledgments}
The authors would like to thank Thomas Brabec, Paul Corkum and Jean-Paul Britcha for many fruitful discussions. This work is supported by NSERC, MRI and CFI.
\end{acknowledgments}

\bibliography{excited_st_XUV_38nm}

\end{document}